\UseRawInputEncoding
\documentclass[aps,prb,twocolumn,showpacs,psfig,superscriptaddress,longbibliography]{revtex4-2}
\usepackage{amsfonts}
\usepackage{mathrsfs}
\usepackage[T1]{fontenc}
\usepackage{ulem}
\usepackage{amsmath}
\usepackage{color}
\usepackage{natbib}
\usepackage{textcomp}
\usepackage{graphicx}
\usepackage{bm}
\usepackage{amssymb}
\usepackage{braket}
\usepackage{xspace}
\usepackage{epstopdf}
\usepackage{dcolumn}
\usepackage{longtable}
\usepackage{multirow}
\usepackage[colorlinks=true, letterpaper=true, pdfstartview=FitV, linkcolor=blue, citecolor=blue, urlcolor=blue]{hyperref}
\usepackage{float}

\makeatletter

\newcommand{\Rmnum}[1]{\expandafter\@slowromancap\romannumeral #1@}
\makeatother

\begin{document}

\title{Superconducting and topological properties of compound Lu$_4$H$_7$N}

 \author{Zheng-Wei Liao}
 \affiliation{School of Physical Sciences, University of Chinese Academy of Sciences, Beijing 100049, China}

 \author{Xin-Wei Yi}
 \affiliation{School of Physical Sciences, University of Chinese Academy of Sciences, Beijing 100049, China}

 \author{Jing-Yang You}
\email{phyjyy@nus.edu.sg}
\affiliation{Department of Physics, National University of Singapore, 2 Science Drive 3, Singapore 117551}

 \author{Bo Gu}
 \email{gubo@ucas.ac.cn}
 \affiliation{Kavli Institute for Theoretical Sciences, and CAS Center for Excellence in Topological Quantum Computation, University of Chinese Academy of Sciences, Beijing 100190, China}

 \author{Gang Su}
 \email{gsu@ucas.ac.cn}
 \affiliation{School of Physical Sciences, University of Chinese Academy of Sciences, Beijing 100049, China}
 \affiliation{Kavli Institute for Theoretical Sciences, and CAS Center for Excellence in Topological Quantum Computation, University of Chinese Academy of Sciences, Beijing 100190, China}

\begin{abstract}
A recent experiment has reported a nitrogen-doped lutetium hydride acheving a remarkable Tc of 294 K at just 1 GPa, significantly reducing the required pressure for obtaining room temperature superconductivity. However, subsequent experimental and theoretical investigations have encountered difficulties in replicating these results, leaving the structure of this Lu-H-N compound shrouded in uncertainty. Here, we propose a stable structure for Lu$_4$H$_7$N employing first-principles calculations. Our calculations reveal that Lu$_4$H$_7$N has a Tc of 1.044 K, which can be substantially enhanced to 11.721 K at 150 GPa, due to the increasing electron-phonon coupling (EPC). Notably, we delve into the nontrivial Z$_2$ band topology of Lu$_4$H$_7$N, featuring discernible surface states near the Fermi level, and we explore its spin Hall conductivity characteristics.
Furthermore, we find that the electron doping can enhance the EPC strength and Tc of Lu$_4$H$_7$N, such as the Lu$_4$H$_7$O structure we predict simulating electron doping for Lu$_4$H$_7$N with an impressive Tc of 3.837 K. This work demonstrates the coexistence of superconducting and topological properties in a Lu-H-N system compound, which holds the promise of guiding the search for novel topological superconducting materials.
\end{abstract}
\maketitle


\section{INTRODUCTION}
Superconducting materials hold immense promise and practical applications across diverse fields, including energy, transportation, and healthcare, thanks to their excellent properties of zero electrical resistance and perfect diamagnetism. Nonetheless, the vast majority of superconductors exhibit disappointingly low superconducting temperatures (Tc), limiting their utility in real-world scenarios. Hence, the search for superconducting materials with substantially higher Tc, ideally at or near room-temperature, has been a longstanding aspiration~\cite{Heikkilae2019}. During the final two decades of the preceding century, a pivotal breakthrough occurred with the discovery of copper-based superconductors capable of achieving Tc in the range of liquid nitrogen temperatures under ambient pressure with a record-high Tc of 130 K reported for the HgBa$_2$Ca$_2$Cu$_3$O$_{8+x}$ structure~\cite{Schilling1993}. According to the tenets of BCS theory, attaining high Tc necessitates a high Debye temperature, which, in turn, demands high vibrational frequency and low atomic mass~\cite{Bardeen1957}. Naturally, hydrogen, characterized by its minuscule atomic mass, emerged as the focus of attention. However, a significant challenge arose from hydrogen's gaseous nature in standard environmental conditions, requiring extreme external pressures to induce metallic behavior. In 2004, Ashcroft proposed that the introduction of other elements surrounding hydrogen atoms can create a chemical pre-compressing environment that could reduce the external pressure required for hydrogen's metallization, which may facilitate the development of high Tc superconductors~\cite{Ashcroft2004}. Since then, many hydrogen-rich materials with high Tc have been predicted, including sulfur hydrides~\cite{Duan2014,Drozdov2015}, rare earth hydrides~\cite{Peng2017,Drozdov2019,Errea2020}, alkali metal hydrides~\cite{Zurek2019}, alkali earth metal hydrides~\cite{Zurek2016}, and transition metal hydrides~\cite{Gao2021}. Among them, the LaH$_{10}$ structure has been synthesized with the highest Tc, reaching 250-260 K at 170-190 GPa, remarkably close to room-temperature~\cite{Drozdov2019,Somayazulu2019}. However, it is evident that despite the impressive progress in enhancing the Tc of hydride superconductors, the critical challenge remains the exceptionally high pressure conditions required, imposing severe limitations on the practical applications of these materials. Therefore, the pursuit of methods to reduce the necessary pressure has emerged as a vital research direction in the realm of hydride superconductors.

Recently, Dasenbrock-Gammon $et~al.$ claimed that they discovered a Lu-H-N system achieving a Tc of 294 K at a low pressure of 1 GPa, much lower than the pressures required for H$_3$S and LaH$_{10}$~\cite{DasenbrockGammon2023}. This groundbreaking finding has sparked substantial interest in the Lu-H-N system as a potential avenue for high-temperature superconductivity. Their hypothesis revolves around the possibility of this compound adopting a Lu$_4$H$_{11}$N structure, wherein a nitrogen atom replaces the hydrogen atom in either octahedral or tetrahedral interstitial sites of the $Fm\overline{3}m$ LuH$_3$ lattice. Unfortunately, many attempts to synthesize Lu-H-N compounds in experiments have not yielded Tc above 1.5 K within the pressure range from ambient to 50.5 GPa~\cite{Shan2023,Zhao2023,Wang2023,Ming2023,Zhang2023,Xing2023,zhang2023electronic,Cai2023,Moulding2023,Li2023a,Liu2023}. In addition, numerous theoretical works have sought to explore various Lu-H-N structures and elucidate the reasons behind the observed color changes in Lu-H-N compounds during experiments~\cite{Liu2023a,Huo2023,Xie2023,Hilleke2023,Ferreira2023,Sun2023,Lucrezi2023,Lu2023,Kim2023,Dangic2023}, but the precise structure remains elusive.

In this article, we found that the Lu$_4$H$_{11}$N structures by replacing the octahedral or tetrahedral interstitial H atom with a N atom respectively and even $Fm\overline{3}m$ LuH$_3$ are dynamically unstable. Therefore, we used CALYPSO~\cite{Wang2010} to search Lu-H-N compounds with radios of Lu:H:N from 4:7:1 to 4:4:4, and obtained a new stable structure Lu$_4$H$_7$N with a space group of $C2/m$ with a Tc of 1.044 K and EPC $\lambda$ of 0.413. By electron doping or applying pressure, the Tc of Lu$_4$H$_7$N can be enhanced. The Tc of Lu$_4$H$_7$O, which is electron doped relative to Lu$_4$H$_7$N, reaches 3.837 K, and at a pressure of 150 GPa, Lu$_4$H$_7$N exhibits a higher Tc of 11.721 K. Notably, Lu$_4$H$_7$N features robust topological properties, including a strong Z$_2$ index and clear surface states near the Fermi level. The spin Hall conductivity (SHC) of Lu$_4$H$_7$N is evaluated to be about 55$\sim$145 $\hbar \cdot (e \cdot \Omega \cdot cm)^{-1}$ at the Fermi level, with the potential to soar to 292 $\hbar \cdot (e \cdot \Omega \cdot cm)^{-1}$ at 0.47 eV.

\section{CALCULATION METHOD}
The first-principles calculations were based on the density-functional theory (DFT) as implemented in the Vienna $ab$ $initio$ (VASP)~\cite{Kresse1996}, $Wannier$90~\cite{Mostofi2014}, $WannierTools$~\cite{Wu2018}, and QUANTUM-ESPRESSO (QE) packages~\cite{Giannozzi2009}, with the projector augmented wave method~\cite{Bloechl1994}. VASP was used to optimize the structure until the forces on atoms are less than 1meV/\AA, and to calculate the electronic band structure, within the Perdew-Burke-Ernzerhof parameterization of the generalized gradient approximation~\cite{Perdew1996}. The plane-wave cutoff energy was taken as 500 eV. The Monkhorst-Pack scheme was used to sample the Brillouin zone with a k-mesh of 9$\times$9$\times$9. 

$Wannier$90 and $Wanniertools$ packages were used to obtain the effective tight-binding Hamiltonian, SHC, surface spectra and topological properties. SHC tensor is calculated by employing the Kubo formula~\cite{Guo2005,Sinova2004}:
\begin{equation}
 \sigma_{\alpha\beta}^{\gamma}=e\hbar\int\frac{dk}{(2\pi)^3}\Omega_{\alpha\beta}^{\gamma}(k),
\end{equation}
where $\Omega_{\alpha\beta}^{\gamma}(k)$ is the k-resolved term obtained by integrating the spin Berry curvature $\Omega_{n,\alpha\beta}^{\gamma}(k)$ of all occupied bands. The \textbf{k}-point mesh for the spin Berry curvature integral adopts 100$\times$100$\times$100, and an extra 5$\times$5$\times$5 fine mesh around those points with $\Omega_{n,\alpha\beta}^{\gamma}(k)$ exceeding 100~\AA$^2$ is added. Spin-orbit coupling (SOC) is considered in SHC calculation.

The calculations of phonon spectra and superconducting properties were performed using QE within density functional perturbation theory (DFPT)~\cite{Baroni2001}. The plane-waves kinetic-energy cutoff was set as 90 Ry and the \textbf{q}-point mesh in the first BZ was taken as 4$\times$4$\times$4. An unshifted \textbf{k}-point mesh of 16$\times$16$\times$16 was utilized.
In the DFPT calculations, Methfessel-paxton smearing method~\cite{Methfessel1989} with a smearing value 0.010 Ry was taken.
To estimate Tc, we use the McMillan-Allen-Dynes equation~\cite{Allen1975,McMillan1968}:
\begin{equation}
	{\rm Tc}=\frac{\omega_{log}}{1.2}{\rm exp}[{-\frac{1.04(1+\lambda)}{\lambda-\mu^*(1+0.62\lambda)}}],
\end{equation}
where $\omega_{log}$ is the logarithmically averaged phonon frequency, and $\lambda$ is a dimensionless parameter describing the EPC strength. The effective screened Coulomb repulsion constant $\mu^{\star}$ was taken as a typical value of 0.1.

\begin{figure}[!htbp]
	\centering
	\includegraphics[scale=0.44,angle=0]{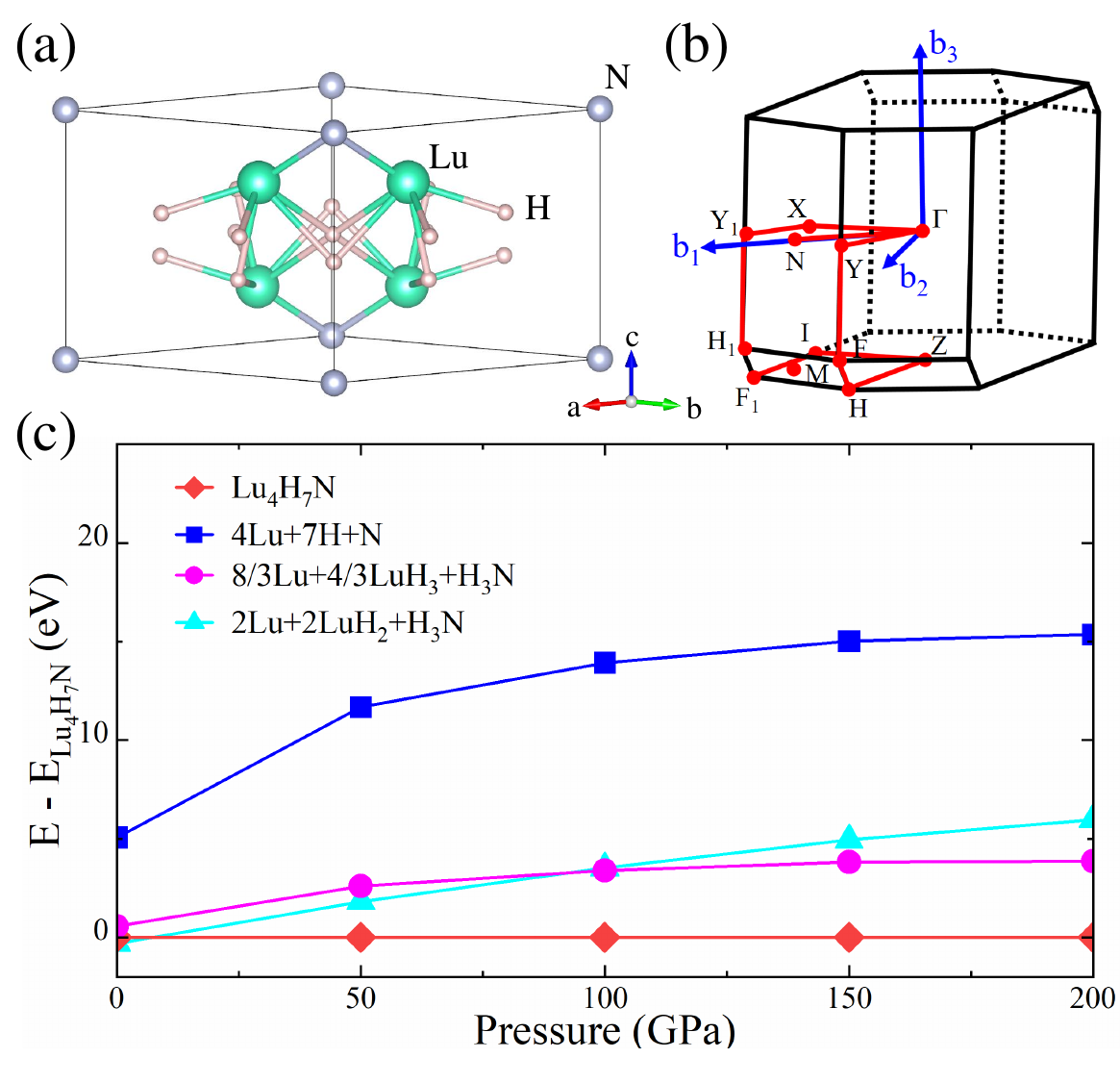}\\
	\caption{(a) The crystal structure, (b) Brillouin zone (BZ), and (c) formation enthalpy of Lu$_4$H$_7$N.}\label{fig1}
\end{figure}

\section{Results}

\subsection{Crystal structure and stability of Lu$_4$H$_7$N}

Given the prevailing ambiguity surrounding the structural configuration of the Lu-H-N system, we adopted an approach by replacing the hydrogen atom in octahedral and tetrahedral interstitial sites of $Fm\overline{3}m$ LuH$_3$ with a nitrogen atom as suggested by Dasenbrock-Gammon $et~al.$, respectively~\cite{DasenbrockGammon2023}. However, upon conducting thorough calculations, we observed that the phonon spectra for these two structures have imaginary frequencies, indicative of dynamic instability. Notably, even LuH$_3$ also displays imaginary phonons (detailed analysis is given in Sec. S1 of the Supplemental Materials~\cite{SuplMat}). Therefore, we focus our exploration on the well-established and stable $Fm\overline{3}m$ LuH$_2$ structure. Employing CALYPSO, we conducted an extensive search for potential structures, varying the Lu:H:N ratio from 4:7:1 to 4:4:4, and obtained a dynamically stable structure Lu$_4$H$_7$N with a $C2/m$ space group as depicted in Fig.$\ref{fig1}$(a). The optimized lattice constants are $a = 6.224$ \AA ~and $c = 5.555$ \AA ~with lattice angles of $\alpha = 91.739^{\circ}$ and $\gamma = 141.526^{\circ}$. In Fig.$\ref{fig1}$(b), we present the Brillouin zone, highlighting the high-symmetry paths $\Gamma$-$Y$-$F$-$H$-$Z$-$I$-$F_1$|$H_1$-$Y$-$X$-$\Gamma$-$N$|$M$-$\Gamma$. In Fig.$\ref{fig1}$(c), we determined the formation enthalpies of Lu$_4$H$_7$N at different pressures to compare with those of multiple synthesis pathways using some synthesized structures, such as Lu ($P6_3/mmc$), H$_2$ ($P6_3/mmc$), N$_2$ ($Pa\overline{3}$), LuH$_3$ ($P\overline{3}c1$), LuH$_2$ ($Fm\overline{3}m$), LuN ($Fm\overline{3}m$), and H$_3$N($P2\uline{~}1\overline{3}$). It can be seen that Lu$_4$H$_7$N boasts a lower energy. 
Furthermore, it's noteworthy that the formation energy of Lu$_2$N$_3$, a compound synthesized experimentally with a space group of $Ia\overline{3}$, stands at 0.86 eV/atom above the hull~\cite{Jain2013,Kieffer1972}. This value is notably higher than the formation energy of Lu$_4$H$_7$N at ambient pressure, which registers at 0.43 eV/atom. Thus, it is possible to obtain this Lu$_4$H$_7$N structure in experiment.

\begin{figure}[!htbp]
	\centering
	\includegraphics[scale=0.48,angle=0]{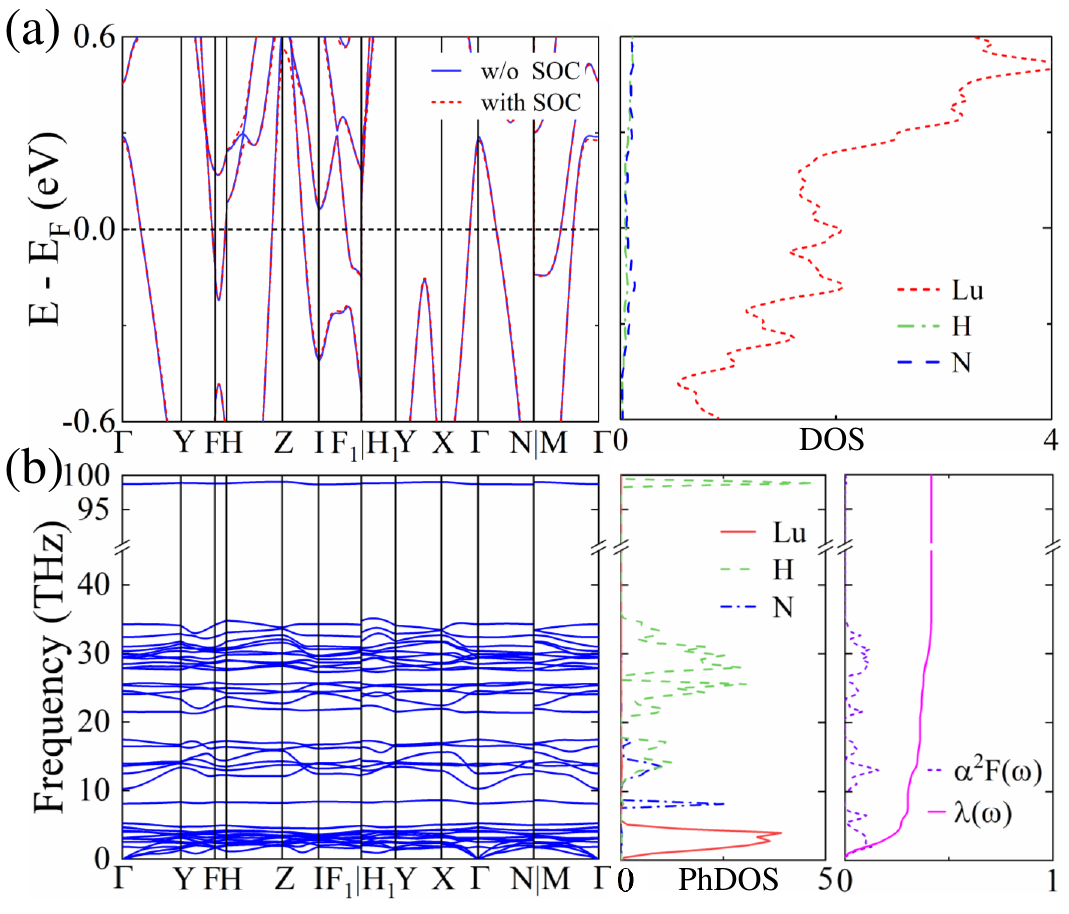}\\
	\caption{(a) The electronic band structure, projected density of states (DOS) without and with spin-orbit coupling (SOC), (b) phonon spectrum, projected phonon density of states (PhDOS), Eliashberg spectral function $\alpha^2F(\omega)$, and cumulative frequency-dependent EPC $\lambda(\omega)$ of Lu$_4$H$_7$N.}\label{fig2}
\end{figure}

\subsection{Superconductivity of Lu$_4$H$_7$N}
Through the first-principles calculations, we find that the Tc of Lu$_4$H$_7$N is 1.044 K with the EPC $\lambda$ 0.413. Fig.$\ref{fig2}$(a) gives the electronic band structures and projected density of states (DOS) for Lu$_4$H$_7$N without and with SOC, revealing its metallic nature. In particular, Lu atoms contribute to the majority of DOS near the Fermi level. In Fig.$\ref{fig2}$(b), the phonon spectra, projected phonon density of states (PhDOS), Eliashberg spectral function $\alpha^2F(\omega)$, and cumulative frequency-dependent EPC $\lambda(\omega)$ of Lu$_4$H$_7$N are comprehensively analyzed. It's evident that vibration modes below 5 THz predominantly originate from Lu atoms, while a distinct flat band around 8THz results from the vibration of N atoms. On the other hand, H atoms play a substantial role in contributing vibration modes higher than 10 THz. Remarkably, the low-frequency vibration region below 10 THz accounts for about 73\% of the total EPC, underscoring the pivotal significance of Lu and N atoms as the principal contributors to the EPC in Lu$_4$H$_7$N.

\begin{figure}[!htbp]
	\centering
	\includegraphics[scale=0.44,angle=0]{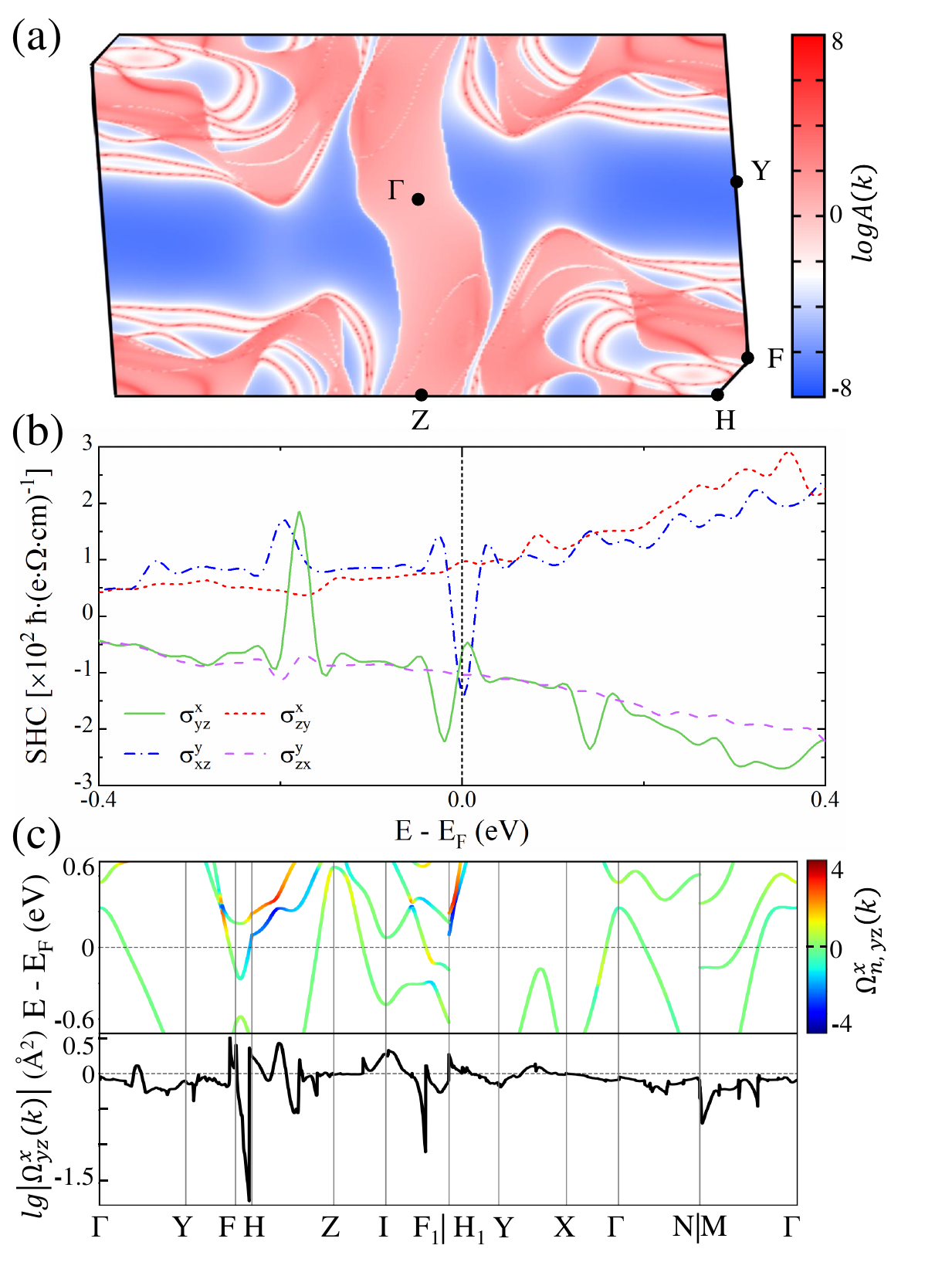}\\
	\caption{(a) The isoenergy slice at 0.2 eV of the surface state spectrum $A(\boldsymbol{k},\omega)$ projected along [010] direction, (b) four larger independent components of spin Hall conductivity (SHC) tensor as a function of chemical potential, (c) band structure weighted by spin Berry curvature $\Omega_{n,zy}^{x}(\boldsymbol{k})$ and k-resolved $\Omega_{yz}^{x}(\boldsymbol{k})$ by integrating the spin Berry curvature of all occupied bands along the high-symmetry paths at the Fermi level for Lu$_4$H$_7$N.}\label{fig3}
\end{figure}

\subsection{Topological properties of Lu$_4$H$_7$N}
Topological superconductivity is a promising method for quantum computing, so the quest for materials exhibiting both superconducting and topological properties has become a focal point of intense interest~\cite{Dong2022,You2022,Yi2022}.
As shown in Fig.~\ref{fig2}(a), several band gapped points emerge along the $H$-$Z$-$I$ paths at around 0.2$\sim$0.5 eV close to the Fermi level because of SOC. To delve deeper into the topological properties, we calculated the Z$_2$ index, yielding a result of (1,001), a clear indication that Lu$_4$H$_7$N qualifies as a topological semimetal~\cite{Fu2007}.
Considering the presence of a mirror symmetry within the $C2/m$ space group, we examined the mirror plane (010) of Lu$_4$H$_7$N. Fig.~\ref{fig3}(c) illustrates an isoenergy slice at 0.2 eV for the surface state spectrum $A(\boldsymbol{k},\omega)$ projected along the [010] direction, which clearly outlines the surface states.

We further investigate the SHC of Lu$_4$H$_7$N from the intrinsic contribution, an intrinsic property contingent upon the electronic band structure and topological characteristics. The crystal symmetry of Lu$_4$H$_7$N imposes constrains on the SHC tensor, encompassing thirteen independent components~\cite{Seemann2015}. We have calculated these thirteen independent components and displayed the four most substantial components $\sigma_{yz}^{x}$, $\sigma_{zy}^{x}$, $\sigma_{xz}^{y}$, and $\sigma_{zx}^{y}$ in Fig.~\ref{fig3}(b). The components exhibit magnitudes in the range of 55$\sim$145 $\hbar\cdot(e\cdot\Omega\cdot cm)^{-1}$ at the Fermi level, with $\sigma_{zy}^{x}$ notably reaching an impressive 292 $\hbar \cdot (e \cdot \Omega \cdot cm)^{-1}$ at 0.47 eV. To reveal the nature of SHC, we plot the band structure of Lu$_4$H$_7$N colored by the magnitude of spin Berry curvature $\Omega_{n,yz}^{x}(\boldsymbol{k})$, as well as the k-resolved $\Omega_{yz}^{x}$ at the Fermi level in Fig.~\ref{fig3}(c). It is obvious that $\Omega_{n,yz}^{x}(\boldsymbol{k})$ shows conspicous peaks aligning with positions of small gapped points, signifying their primary contributions to SHC~\cite{You2023}.

\begin{table}[htbp]
	\renewcommand\arraystretch{1.25}
	\caption{The DOS at the Fermi level $N(E_F)$ (in unit of states/spin/eV/cell), volume of primitive cell $V$ (\AA$^3$), $\omega_{log}$ (K), $\lambda$, and Tc (K) for Lu$_4$H$_7$N at ambient pressure, 10, 100, 150, and 200 GPa.}
	\label{T-1}
	\begin{tabular}{l<{\centering}p{1.3cm}<{\centering}p{1.2cm}<{\centering}p{1.2cm}<{\centering}p{1.2cm}<{\centering}p{1.2cm}<{\centering}p{1.3cm}<{\centering}}
		\hline 
		\hline
		&$P$(GPa)   &$N(E_F$) &$V$(\AA$^3$)  &$\omega_{log}$(K)   &$\lambda$ &Tc(K) \\
		\hline 
		&0      &18.53  &134.13   &207.3     &0.413     &1.044          \\
		&10     &17.88  &120.12   &213.0     &0.464     &1.892         \\
		&100    &13.10  &77.28   &214.5     &0.724     &8.097         \\
		&150    &12.18  &69.31   &219.4     &0.860     &11.721         \\
		&200    &11.66  &64.09   &237.4     &0.773     &10.355          \\
		\hline \hline
	\end{tabular}
\end{table}

\begin{table}[htbp]
	\renewcommand\arraystretch{1.25}
	\caption{The DOS at Fermi level $N(E_F)$ (in unit of states/spin/eV/cell), volume of primitive cell $V$ (\AA$^3$), $\omega_{log}$ (K), $\lambda$, and Tc (K) for Lu$_4$H$_7$N, LuH$_2$, Lu$_4$H$_7$C and Lu$_4$H$_7$O at ambient pressure.}
	\label{T-2}
	\begin{tabular}{l<{\centering}p{1.3cm}<{\centering}p{1.3cm}<{\centering}p{1.3cm}<{\centering}p{1.3cm}<{\centering}p{1.3cm}<{\centering}p{1.2cm}<{\centering}}
		\hline 
		\hline
		&$N(E_F$) &$V$(\AA$^3$)  &$\omega_{log}$(K)   &$\lambda$ &Tc(K) \\
		\hline 
		\multirow{1}{*}{Lu$_4$H$_7$N}
		&18.53  &134.13   &207.3     &0.413     &1.044          \\
		\multirow{1}{*}{LuH$_2$}
		&3.95  &31.67   &297.4     &0.258     &0.024          \\
		\multirow{1}{*}{Lu$_4$H$_7$C}
		&8.32  &136.88   &232.6     &0.235     &0.059         \\
		\multirow{1}{*}{Lu$_4$H$_7$O}
		&20.89  &134.81   &149.7     &0.624     &3.837         \\
		\hline \hline
	\end{tabular}
\end{table}

\begin{figure}[!htbp]
	\centering
	\includegraphics[scale=0.48,angle=0]{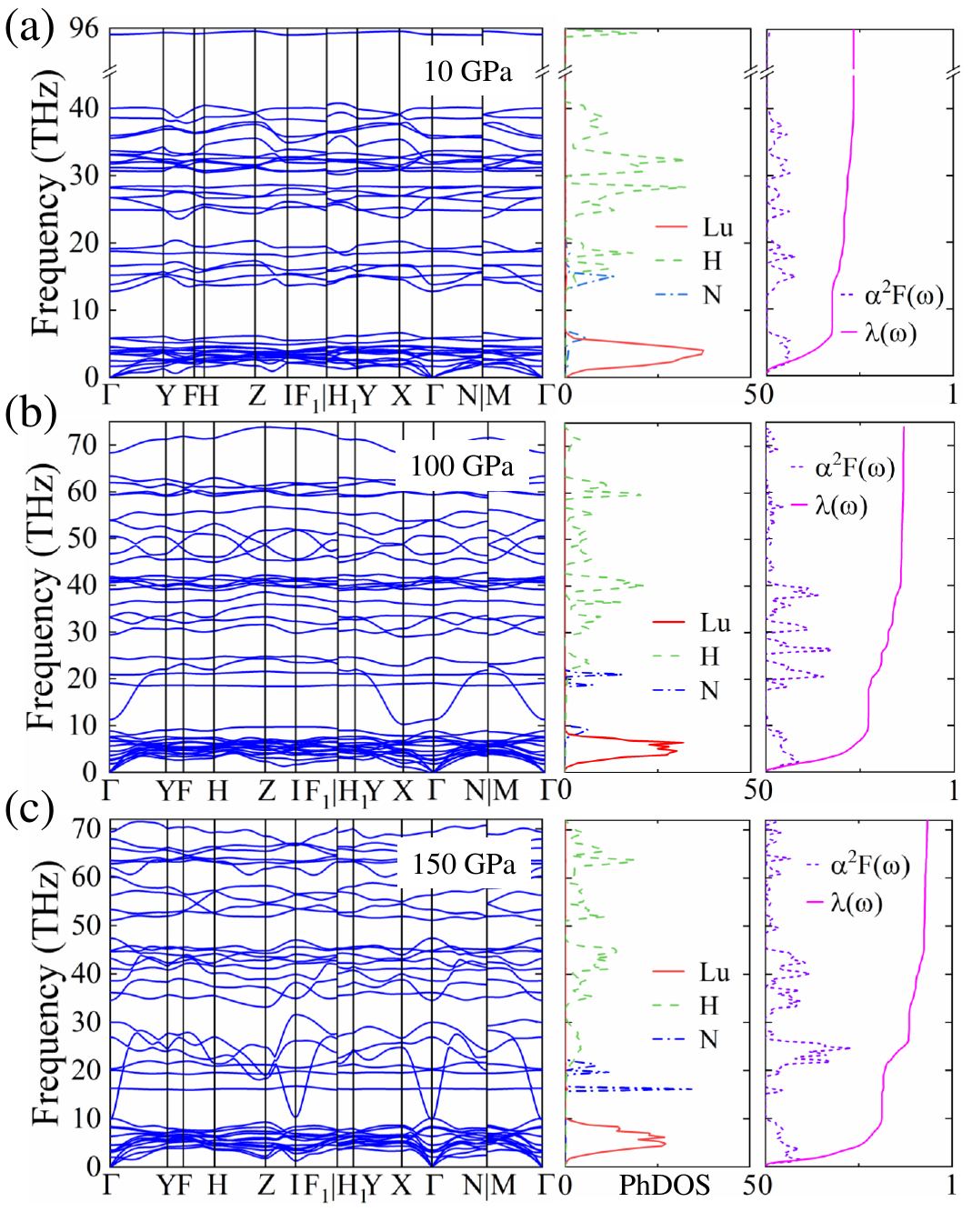}\\
	\caption{The phonon spectra, projected PhDOS, $\alpha^2F(\omega)$, and $\lambda(\omega)$ of Lu$_4$H$_7$N at (a) 10, (b) 100, and (c) 150 GPa.}\label{fig4}
\end{figure}

\begin{figure}[!htbp]
	\centering
	\includegraphics[scale=0.48,angle=0]{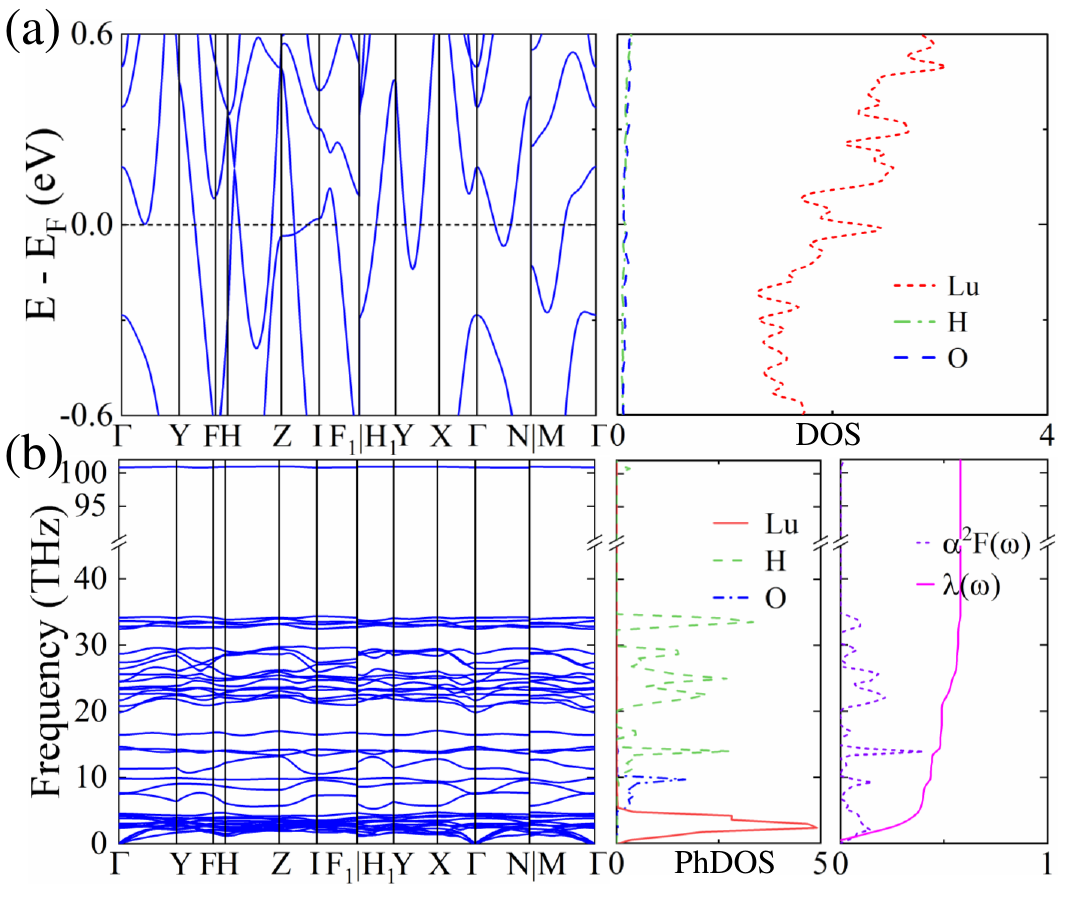}\\
	\caption{(a) The electronic band structure, projected DOS, (b) phonon spectrum, projected PhDOS, $\alpha^2F(\omega)$, and $\lambda(\omega)$ of Lu$_4$H$_7$O.}\label{fig5}
\end{figure}

\subsection{Discussion}
To gain a deeper insight into the superconducting properties of Lu$_4$H$_7$N, we conducted calculations in two aspects. First, we applied pressure to Lu$_4$H$_7$N and found that upon reaching a pressure of 10 GPa, the structure remained dynamically stable, and the Tc reaches 1.892 K. However, with continued pressure increasing, the volume of Lu$_4$H$_7$N decreases, causing the N atoms to draw nearer to the surrounding Lu atoms which intensifies the repulsive forces, ultimately leading to dynamic instability. At an elevated pressure of 100 GPa, the positions of N atoms changed, resulting in the structure regaining its dynamical stability (for a detailed analysis, please refer to Sec. S3 of the Supplemental Materials~\cite{SuplMat}). Therefore, we postulate the existence of another structural form of Lu$_4$H$_7$N in the pressure range between 10 and 100 GPa. For the stable Lu$_4$H$_7$N structure above 100 GPa, we calculated its superconducting properties, as meticulously detailed in Table~\ref{T-1}. As the pressure increases, the electron DOS at the Fermi level gradually decreases, while $\omega_{log}$ steadily increases. As observed, the Tc of Lu$_4$H$_7$N reaches 11.721 K, with an enhanced EPC $\lambda$ of 0.86 at 150 GPa. Fig.~\ref{fig4} presents the superconducting properties of Lu$_4$H$_7$N at 10, 100, and 150 GPa. Compared to Lu$_4$H$_7$N at ambient pressure and near-ambient pressure (10 GPa), it becomes evident that at pressures exceeding 100 GPa, the vibrations originating from Lu atoms, which constitute the primary contribution to $\lambda$, extend from the 0$\sim$5 THz range to 0$\sim$10 THz,, significantly increasing $\lambda$ and driving an increase of Tc.

In addition, we replaced the N atom in Lu$_4$H$_7$N with a C or O atom to simulate hole or electron doping, respectively. Subsequently, we conducted thorough assessments of their superconducting properties, and proceeded to compare them with the well-known $Fm\overline{3}m$ LuH$_2$, as outlined in Table~\ref{T-2}. The electron DOS of LuH$_2$ and Lu$_4$H$_7$C at the Fermi level are much lower than that of Lu$_4$H$_7$N, weakening the EPC strength with a Tc less than 0.1 K. In stark contrast, as shown in Fig.~\ref{fig5}, we observe a higher electron DOS at the Fermi level and PhDOS of Lu atoms in Lu$_4$H$_7$O, which serve as the primary source of $\lambda$, consequently enhancing the EPC strength and yielding an impressive Tc of 3.837 K.


\section{SUMMARY}
To summarize, we predicted a new stable Lu-H-N compound Lu$_4$H$_7$N, which has a Tc of 1.044 K. Additionally, Lu$_4$H$_7$N exhibits a strong Z$_2$ index and clear surface states near the Fermi level. The SOC opens several band gaps near the Fermi level, where large spin Berry curvatures exist and contribute significantly to the SHC of Lu$_4$H$_7$N. Furthermore, we found that the Tc of Lu$_4$H$_7$N can be enhanced to 11.721 K at 150 GPa. 
We also performed element substitution based on the Lu$_4$H$_7$N structure to simulate hole and electron doping and obtained two stable materials Lu$_4$H$_7$C and Lu$_4$H$_7$O with calculated Tc of 0.059 and 3.837 K, respectively. The change in Tc is primarily attributed to the modification of electronic DOS following doping.
Our study reveals the coexistence of superconducting and topological properties within the Lu$_4$H$_7$N structure, making it valuable for advancing research on novel topological superconducting materials.


\acknowledgements
This work is supported in part by the Innovation Program for Quantum Science and Technology (No. 2021ZD0301800), the National Key R\&D Program of China (Grant No. 2018YFA0305800), the Strategic Priority Research Program of the Chinese Academy of Sciences (Grant No. XDB28000000), the National Natural Science Foundation of China (Grant No.11834014), and Beijing Municipal Science and Technology Commission (Grant No. Z118100004218001). B.G. is supported in part by the National Natural Science Foundation of China (Grant No. 12074378), the Chinese Academy of Sciences (Grant No. YSBR-030), the Strategic Priority Research Program of Chinese Academy of Sciences (Grant No. XDB33000000), and the Beijing Natural Science Foundation（(Grant No. Z190011).


%

\end{document}